\documentclass[pra,aps,showpacs,showkeys,superscriptaddress,twocolumn]{revtex4}

\usepackage{graphicx}

\begin{document}

\title{Charge Qubit Storage and its Engineered Decoherence via Microwave
Cavity}
\author{Y.B. Gao}
\email{gaoyb@itp.ac.cn}
\affiliation{Applied Physics Department,
Beijing University of Technology, Beijing, 100022, China}
\affiliation{Institute of Theoretical Physics, Chinese Academy of
Sciences, Beijing, 100080, China}
\author{C. Li}
\affiliation{Institute of Theoretical Physics, Chinese Academy of Sciences, Beijing,
100080, China}
\date{\today}

\begin{abstract}
We study the entanglement of the superconducting charge qubit with the
quantized electromagnetic field in a microwave cavity. It can be controlled
dynamically by a classical external field threading the SQUID within the
charge qubit. Utilizing the controllable quantum entanglement, we can
demonstrate the dynamic process of the quantum storage of information
carried by charge qubit. On the other hand, based on this engineered quantum
entanglement, we can also demonstrate a progressive decoherence of charge
qubit with quantum jump due to the coupling with the cavity field in
quasi-classical state.
\end{abstract}

\maketitle

\section{introduction}

The concept of quantum entanglement is crucial to the fundamental aspects of
quantum mechanics, including the quantum measurement and quantum decoherence
\cite{Zurek83}. It also dominates the development of quantum information
science and technology. At this stage, implementing the realistic quantum
computation and communication, one needs to create the maximally quantum
entanglement for potentially physical carrier of quantum information.

As potential candidates of information carrier in quantum computation, some
kinds of qubit based on Josephson junctions \cite{Makhlin01}, such as charge
qubit, phase qubit and flux qubit, have been implemented experimentally. Now
single qubit operation and two qubit logic gate based on Josephson junctions
demonstrate an important role to realize a scalable quantum computing.
Recent experiments demonstrate the Rabi oscillation in a Cooper-pair box
(charge qubit) \cite{Nakamura99} and indicate existence of entangled
two-qubit states \cite{Pashkin03}. Up to now, the decoherence time which is
of order $5\mu s$ has been reported in \cite{Han02}. In the process of
quantum computing, we need qubits with long decoherence time, so we must
find a longer life time medium to store the state of qubit. The microwave
cavity is a good candidate which life time is of order $1ms$ \cite{Raimond01}%
. Actually quantum information storage is the central issues in the study of
quantum information and computation \cite{Lukin00,Lukin03,Li03,Cleland03}.

In this paper, we describe a model of a charge qubit coupled to a single
mode quantum cavity field. Under the rotating wave approximation (RWA), we
demonstrate how to implement a dynamic process of quantum information
storage in the qubit-cavity system. In addition, we consider the
entanglement and engineered decoherence of the qubit-cavity system.

In fact, the integration of qubit and cavity QED \cite{Han03,You03,Girvin03}
has become the focus of quantum information storage and quantum computation.
In order to investigate quantum coherence and information storage in the
qubit-cavity system, we consider a single mode quantum cavity field with
frequency $\omega \sim 30GHz$ (typically in the microwave domain) coupled to
a charge qubit (with dc SQUID) in which the charging energy $E_{C}\sim
122\mu eV$ and the Josephson coupling energy $E_{J}\sim 34\mu eV$ \cite%
{Nakamura02}.

\begin{figure}[ht]
\includegraphics[width=8cm]{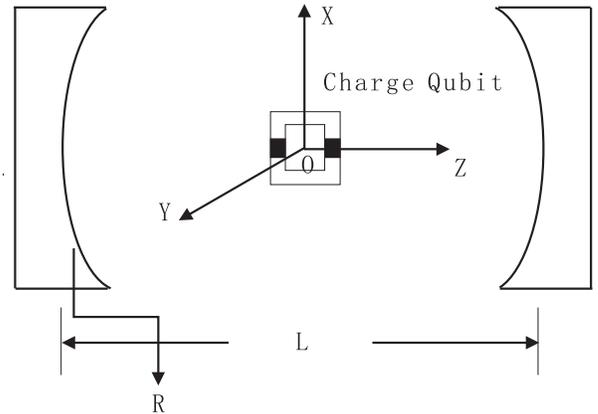}
\caption{Schematic of the charge qubit-cavity system. Superconducting
microwave cavity with parameter, R=2.55mm, L=0.5cm. }
\label{cavity}
\end{figure}

\section{charge qubit coupled to cavity field}

The charge qubit considered in this paper is a dc SQUID consisting of two
identical Josephson junctions enclosed by a superconducting loop. The
Hamiltonian for an dc SQUID can be written as in \cite{Makhlin99}
\begin{equation}
H=4E_{C}\left( n_{g}-\frac{1}{2}\right) \sigma _{z}-E_{J}\cos \left( \pi
\frac{\Phi }{\Phi _{0}}\right) \sigma _{x}+\omega a^{\dagger }a  \label{Ham1}
\end{equation}%
where $E_{C}$ is the charging energy, $E_{J}$ Josephson coupling energy and
we assume $E_{C}>>E_{J}$, and $\Phi $ the magnetic flux generated by
controlled classical magnetic field and quantum cavity field. Quasi-spin
operators
\[
\sigma _{z}=\left\vert 0\right\rangle _{q}\left\langle 0\right\vert
_{q}-\left\vert 1\right\rangle _{q}\left\langle 1\right\vert _{q},\sigma
_{x}=\left\vert 0\right\rangle _{q}\left\langle 1\right\vert _{q}+\left\vert
1\right\rangle _{q}\left\langle 0\right\vert _{q}
\]%
are defined in the charge qubit basis ($\left\vert 0\right\rangle _{q}$ and $%
\left\vert 1\right\rangle _{q}$), $\Phi _{0}=\frac{h}{2e}$ denotes the flux
quanta.

Now we consider the case that the magnetic flux threading the dc SQUID is
generated by magnetic field $\mathbf{B}$ consisting of external classical
magnetic field $\mathbf{B}_{e}$ and quantum cavity field $\mathbf{B}_{f}$,
i.e., \cite{You03}
\[
\mathbf{B}=\mathbf{B}_{e}+\mathbf{B}_{f}.
\]
So the total magnetic flux threading SQUID is also divided into two parts%
\[
\Phi =\Phi _{e}+\Phi _{f},
\]
where $\Phi _{e}=\int \mathbf{B}_{e}\cdot d\mathbf{S}$ is the external
classical flux threading the dc SQUID, $\Phi _{f}=\int \mathbf{B}_{f}\cdot d%
\mathbf{S}$ the cavity-induced quantum flux through the dc SQUID and $S$ any
area bounded by the dc SQUID.

We assume that there exists a single mode standing wave cavity field of
frequency $\omega $ \cite{Scully97}. The quantum cavity field is assumed to
be transverse with respect to the magnetic field $\mathbf{B}_{f}$ polarized
in the y-direction, i.e.,
\begin{equation}
B_{y}(z)=-i\left( \frac{\hbar \omega }{\varepsilon _{0}Vc^{2}}\right) ^{%
\frac{1}{2}}\left( a-a^{\dagger }\right) \cos \left( \frac{2\pi }{\lambda }%
z\right)  \label{m-field}
\end{equation}%
where $a^{\dagger }(a)$ is the creation (annihilation) operator of the
single mode cavity field and $V$ the electromagnetic mode volume in cavity.

In Fig.\ref{cavity}, the dc SQUID lies in the x-z plane and at the position
of the antinode of standing wave field, i.e., $z=0$. Then the magnetic field
$B_{y}(z)$ will be

\[
B_{y}(0)=-i\left( \frac{\hbar \omega }{\varepsilon _{0}Vc^{2}}\right) ^{%
\frac{1}{2}}\left( a-a^{\dagger }\right)
\]%
and magnetic flux $\Phi _{f}$ is given by
\begin{equation}
\Phi _{f}=-i\left( \frac{\hbar \omega }{\varepsilon _{0}Vc^{2}}\right) ^{%
\frac{1}{2}}S\left( a-a^{\dagger }\right) .  \label{q-flux}
\end{equation}%
In this case, we can derive the spin-Boson Hamiltonian from Eq.(\ref{Ham1})
in a straightforward way,
\begin{equation}
H=4E_{C}\left( n_{g}-\frac{1}{2}\right) \sigma _{z}-E_{J}\cos \left( \phi
_{e}+\phi _{f}\right) \sigma _{x}+\omega a^{\dagger }a  \label{Ham5}
\end{equation}%
where
\[
\phi _{e}=\frac{\pi \Phi _{e}}{\Phi _{0}},\phi _{f}=-i\phi _{0}\left(
a-a^{\dagger }\right)
\]%
and
\[
\phi _{0}=\frac{\pi S}{\Phi _{0}}\left( \frac{\hbar \omega }{\varepsilon
_{0}Vc^{2}}\right) ^{\frac{1}{2}}.
\]

As shown in Fig.\ref{cavity}, two spherical mirrors form microwave cavity
containing a single mode standing wave field and an external classical
magnetic field is also injected into the cavity. In this paper, we adopt
some appropriate parameters for the geometry of cavity, the curvature radius
$R=2\,.55$mm, the width between two mirrors $L=0.5$cm. By some simple
calculations, we get that the cavity field $B=\left( \frac{\hbar \omega }{
\epsilon _{0}Vc^{2}}\right) ^{\frac{1}{2}}=7.52\times 10^{-11}$(Tesla) and $%
\phi _{0}=\pi \frac{\Phi _{f}}{\Phi _{0}}=1.14\times 10^{-5}$. In a low
photon number cavity, we find that $\phi _{f}\ll \phi _{e}$, thus there is
only a weak polynomial nonlinearity in Eq.(\ref{Ham5}).

\section{dynamical quantum information-storage of charge qubit}

Any quantum computer realized in a laboratory will be subject to the
influence of environmental noise in the preparation, manipulation, and
measurement of quantum-mechanical states which causes quantum decoherence.
In the implementation of quantum computation, quantum information should be
encoded in a quantum network formed by many qubits with longer decoherence
time. Subsequently one can store, manipulate and communicate quantum
information \cite{DiVincenzo00}. So we are interested in study of quantum
information-storage.

Now up to the first order of $\phi _{f}$ in Eq.(\ref{Ham5}), we consider
quantum information-storage in the qubit-cavity system. To this end, we
adopt the appropriate parameters, i.e., the frequency of microwave cavity at
$\omega \sim 30GHz$ and the charging energy of charge qubit at $E_{C}\sim
29.5GHz$ \cite{Nakamura02}. When we tune gate voltage of Josephson junction
at $n_{g}\sim 0.627$, the RWA is valid.

In this section, we consider the storage of quantum information of the
charge qubit in the cavity mode. We derive the Hamiltonian for qubit-cavity
system from Eq.(\ref{Ham5}),
\begin{equation}
H=\frac{\omega }{2}\sigma _{z}+i\eta \left( a^{\dagger }\sigma _{-}-a\sigma
_{+}\right) +\omega a^{\dagger }a.  \label{RWA-Ham}
\end{equation}%
It is just as same as a Jaynes-Cummings model \cite{Jaynes63} in optical
cavity QED.

Now we demonstrate a process of quantum information transfer from one charge
qubit to cavity field. We assume that the initial state of charge qubit is
in an arbitrary state $\alpha \left\vert 0\right\rangle _{q}+\beta
\left\vert 1\right\rangle _{q}$ and we want transfer this state from charge
qubit to cavity that its initial state is in the vacuum state $\left\vert
0\right\rangle _{c}$. The process of information storage from qubit to
cavity can be described by
\begin{equation}
\left( \alpha \left\vert 0\right\rangle _{q}+\beta \left\vert 1\right\rangle
_{q}\right) \otimes \left\vert 0\right\rangle _{c}\rightarrow \left\vert
0\right\rangle _{q}\otimes \left( \alpha \left\vert 0\right\rangle
_{c}+\beta \left\vert 1\right\rangle _{c}\right) .  \label{storage}
\end{equation}

From Eq.(\ref{storage}), we find that this information-storage process could
be done with a transformation \cite{Han03}%
\begin{eqnarray}
\left\vert 0\right\rangle _{q}\left\vert 0\right\rangle _{c} &\rightarrow
&\left\vert 0\right\rangle _{q}\left\vert 0\right\rangle _{c},  \label{rule}
\\
\left\vert 1\right\rangle _{q}\left\vert 0\right\rangle _{c} &\rightarrow
&\left\vert 0\right\rangle _{q}\left\vert 1\right\rangle _{c}.  \nonumber
\end{eqnarray}
We can easily verify that the Hamiltonian in Eq.(\ref{RWA-Ham}) holds the
condition of information storage in Eq.(\ref{rule}), i.e.,
\begin{eqnarray*}
H\left\vert 0\right\rangle _{q}\left\vert 0\right\rangle _{c} &=&-\frac{
\omega }{2}\left\vert 0\right\rangle _{q}\left\vert 0\right\rangle _{c} \\
H\left\vert 1\right\rangle _{q}\left\vert 0\right\rangle _{c} &=&\frac{
\omega }{2}\left\vert 1\right\rangle _{q}\left\vert 0\right\rangle
_{c}+i\eta \left\vert 0\right\rangle _{q}\left\vert 1\right\rangle _{c}
\end{eqnarray*}

Now the process of quantum state transfer from the charge qubit to microwave
cavity can be realized in the following steps.

Firstly at the time $t=0$, we turn off the external classical field and
prepare the initial state of qubit in the pure state
\[
\alpha \left\vert 0\right\rangle _{q}+\beta \left\vert 1\right\rangle _{q}
\]
corresponding to the reduced density matrix of the qubit%
\[
\rho _{q}\left( 0\right) =\left(
\begin{array}{cc}
\left\vert \alpha \right\vert ^{2} & \alpha \beta ^{\ast } \\
\alpha ^{\ast }\beta & \left\vert \beta \right\vert ^{2}%
\end{array}
\right) .
\]
where $\left\vert \alpha \right\vert ^{2}+\left\vert \beta \right\vert
^{2}=1 $, the cavity is prepared in the vacuum state $\left\vert
0\right\rangle _{c}$. The initial state of the qubit-cavity system can be
written as
\begin{equation}
\left\vert \psi (0)\right\rangle =\left( \alpha \left\vert 0\right\rangle
_{q}+\beta \left\vert 1\right\rangle _{q}\right) \otimes \left\vert
0\right\rangle _{c} .  \label{init-storage}
\end{equation}

Secondly with time increasing, we turn on the external classical field and
allow the qubit and microwave cavity to interact on resonance. Then the
state of the qubit-cavity system evolves into
\begin{eqnarray*}
\left\vert \psi \left( t\right) \right\rangle &=&\left\vert 0\right\rangle
_{q}\otimes \left( \alpha e^{i\frac{\omega }{2}t}\left\vert 0\right\rangle
_{c}-\beta e^{-i\frac{1}{2}\omega t}\sin \eta t\left\vert 1\right\rangle
_{c}\right) \\
&&+\beta e^{-i\frac{1}{2}\omega t}\cos \eta t\left\vert 1\right\rangle
_{q}\left\vert 0\right\rangle _{c}
\end{eqnarray*}
corresponding to the reduced density matrix of the cavity
\begin{equation}
\rho _{c}\left( t\right) =\left(
\begin{array}{cc}
\left\vert \alpha \right\vert ^{2}+\left\vert \beta \right\vert ^{2}\cos
^{2}\eta t & -\alpha \beta ^{\ast }e^{i\omega t}\sin \eta t \\
-\alpha ^{\ast }\beta e^{-i\omega t}\sin \eta t & \left\vert \beta
\right\vert ^{2}\sin ^{2}\eta t%
\end{array}
\right).  \label{t-storage}
\end{equation}
Obviously, we find that when $\eta t=0$,
\begin{equation}
\rho _{c}\left( 0\right) =\left(
\begin{array}{cc}
1 & 0 \\
0 & 0%
\end{array}
\right)
\end{equation}
and the transfer of quantum information occurs when $\eta t=\frac{\pi }{2}$
or $t=\frac{\pi }{2\eta }$,
\[
\rho _{c}\left( \frac{\pi }{2\eta }\right) =\left(
\begin{array}{cc}
\left\vert \alpha \right\vert ^{2} & -\alpha \beta ^{\ast }e^{i\frac{\omega
\pi }{2\eta }} \\
-\alpha ^{\ast }\beta e^{-i\frac{\omega \pi }{2\eta }} & \left\vert \beta
\right\vert ^{2}%
\end{array}
\right)
\]
and the state of the qubit-cavity system evolves into the state
\begin{equation}
\left\vert \psi (\frac{\pi }{2\eta })\right\rangle =\left\vert
0\right\rangle _{q}\otimes \left( \alpha e^{i\frac{\omega \pi }{4\eta }
}\left\vert 0\right\rangle _{c}-\beta e^{-i\frac{\omega \pi }{4\eta }
}\left\vert 1\right\rangle _{c}\right) ,  \label{time-storage}
\end{equation}
where the cavity is in the pure state
\[
\alpha e^{i\frac{\omega \pi }{4\eta }}\left\vert 0\right\rangle _{c}-\beta
e^{-i\frac{\omega \pi }{4\eta }}\left\vert 1\right\rangle _{c}.
\]
From Eq.(\ref{init-storage}) and Eq.(\ref{time-storage}), we can say that at
time $t=\frac{\pi }{2\eta }$, the quantum information contained in the qubit
has been stored in the microwave cavity.

\begin{figure}[th]
\includegraphics[width=8cm]{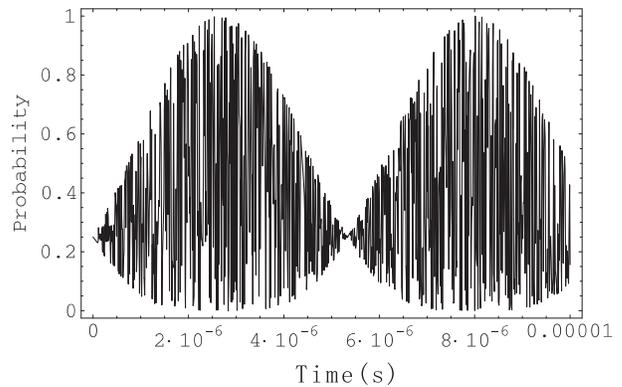}
\caption{The probability of the system in the state $\left\vert \protect\psi %
(\frac{\protect\pi }{2\protect\eta })\right\rangle $. The qubit transfer
quantum information to the microwave cavity when the probability is equal to
1.}
\label{t6}
\end{figure}

As shown in Fig.\ref{t6}, we set $\alpha =\beta =\frac{1}{\sqrt{2}}$ and
demonstrate the probability of that quantum information of the qubit is
transferred to the microwave cavity
\begin{equation}
P=\frac{1}{4}+\frac{1}{2}\cos \left( \omega t-\frac{\omega \pi }{2\eta }%
\right) \sin \eta t+\frac{1}{4}\sin ^{2}\eta t.  \label{trans-P}
\end{equation}%
We find that the qubit transfers the quantum information to the microwave
cavity at the time $t=\frac{\pi }{2\eta }\sim 2.7\mu s$.

\section{entanglement and engineered decoherence}

In previous work on cavity QED \cite{Raimond97}, Raimond et al dealt with
entanglement and decoherence for the cavity-atom system and show a
reversible decoherence process of a mesoscopic superposition of field
states. In this paper, our model is quite similar to a cavity QED model
without the rotation-wave-approximation (RWA), which usually describes the
single mode cavity interacting with an two-level atom \cite{Sun00}. In this
cavity QED model, when the atoms in different states $\left\vert
0\right\rangle $ and $\left\vert 1\right\rangle $ will modify the states of
cavity field in different ways and thus induce the quantum decoherence of
atomic states superposition. In this section, we consider these issues about
entanglement and decoherence in the qubit-cavity system.

Now we consider a single charge qubit (dc SQUID) coupled to a single-mode
cavity. According to Eq.(\ref{Ham5}), we tune the gate voltage $V_{g}$ such
that $n_{g}=\frac{1}{2}$ and then get a Hamiltonian for a standard quantum
measurement model,
\begin{equation}
H=-E_{J}\cos \left( \phi _{e}+\phi _{f}\right) \sigma _{x}+\omega a^{\dagger
}a.  \label{m-Ham}
\end{equation}

By some simple calculations from Eq.(\ref{m-Ham}), we derive an effective
Hamiltonian,
\begin{equation}
H=H_{0}\left\vert 0\right\rangle \left\langle 0\right\vert +H_{1}\left\vert
1\right\rangle \left\langle 1\right\vert  \label{e-Ham}
\end{equation}%
which is diagonal with respect to $\left\vert 0\right\rangle =\left\vert
0\right\rangle _{q}+\left\vert 1\right\rangle _{q}$ and $\left\vert
1\right\rangle =\left\vert 0\right\rangle _{q}-\left\vert 1\right\rangle
_{q} $ are eigenstates of quasi spin operator $\sigma _{x}$. The effective
actions on cavity field are
\begin{equation}
H_{k}=-\left( -1\right) ^{k}E_{J}\cos \left( \phi _{e}+\phi _{f}\right)
\sigma _{x}+\omega a^{\dagger }a  \label{k-Ham}
\end{equation}%
for $k=0,1$ respectively. Obviously the Hamiltonian in Eq.(\ref{e-Ham}) can
create entanglement between charge qubit and cavity field. Since the
different actions on the cavity field are exerted by the charge qubit in
different quasi-spin state.

Technically the engineered decoherence process is described by the time
evolution of the reduced density matrix of the coupled qubit-cavity system.
To analyze it, we can derive the reduced density matrix for the time
evolution of the qubit-cavity system. The decoherence process means that the
off diagonal elements of the reduced density matrix of the qubit vanish,
while the diagonal elements remain unchanged.

Here we get time evolution of density matrix for the qubit-cavity system
\begin{equation}
\rho \left( t\right) =\left\vert \psi \left( t\right) \right\rangle
\left\langle \psi \left( t\right) \right\vert  \label{density}
\end{equation}
and calculate the reduced density matrix of the qubit
\begin{eqnarray}
\rho _{s}\left( t\right) &=&C_{0}^{\ast }C_{0}\left\vert 0\right\rangle
\left\langle 0\right\vert +C_{1}^{\ast }C_{1}\left\vert 1\right\rangle
\left\langle 1\right\vert  \label{reduced density} \\
&&+\left\langle d_{1}\left( t\right) |d_{0}\left( t\right) \right\rangle
C_{1}^{\ast }C_{0}\left\vert 0\right\rangle \left\langle 1\right\vert +h.c
\nonumber
\end{eqnarray}

We usually define decoherence factor as a measure of the coherence of system
\cite{Sun93},
\begin{equation}
D\left( t\right) =\left\vert \left\langle d_{1}\left( t\right) |d_{0}\left(
t\right) \right\rangle \right\vert.  \label{dfactor}
\end{equation}
At time $t=0$, the decoherence factor $D\left( 0\right) =1$, i.e., the
system owns good coherence. As time passes $t>0$, the decoherence factor is
not equal to one, $D\left( t\right) \neq 1$. Due to the coupling between
qubit and cavity, quantum information of the qubit continuously leaks into
the cavity, until the states of the cavity correlated to the different
states of the system become orthogonal, i.e., $D\left( t\right) =0$ and the
quantum information contained in the qubit can be described classically. The
reduced density matrix of the qubit turns continuously into a incoherent
mixture.

Because $\phi _{f}\ll \phi _{e}$, we expand $\cos \left( \phi _{e}+\phi
_{f}\right) $ into a series of $a$ ($a^{\dagger }$) to the second order of $%
\phi _{f}$ and then get
\begin{equation}
H=-\left[ \delta aa-\delta aa^{\dagger }+i\eta a+h.c.+E_{J}\cos \phi _{e}%
\right] \sigma _{x}+\omega a^{\dagger }a  \label{Ham6}
\end{equation}%
where
\[
\delta =\frac{1}{2}\phi _{0}^{2}E_{J}\cos \phi _{e}\sim 10^{-10}E_{J}
\]%
and
\[
\eta =\phi _{0}E_{J}\sin \phi _{e}\sim 10^{-5}E_{J}
\]%
are the first order interaction strength and the second order interaction
strength\ respectively. Clearly the above Hamiltonian includes the
interaction terms of $\eta $, which drives the cavity field into the
coherent state and the interaction with terms of $\delta $ which drives the
cavity field into the squeezed state.

In this paper, we control the interaction between qubit and cavity by
changing the strength of the external classical field, i.e., $\phi _{e}$.
From Eq.(\ref{Ham6}), if we turn off the external field, i.e., $\sin \phi
_{e}=0$, the above Hamiltonian only includes the second order interaction.
If we turn on the external magnetic field, i.e., $\cos \phi _{e}=0$, the
above Hamiltonian only includes the first order interaction.

Now we choose the coherent state $\left\vert \alpha \right\rangle $ as the
initial state of quantum cavity field. It is not difficult to prepare such
quasi-classical state by a driving current source. Through some simple
calculations, we find that the Hamiltonian in Eq.(\ref{Ham6}) containing
terms of $aa$ and $a^{\dagger }a^{\dagger }$ will drive the cavity field
into two different squeezed states
\begin{equation}
\left\vert d_{k}\left( t\right) \right\rangle =\left\vert \beta _{k},\mu
_{k},\nu _{k}\right\rangle e^{i\theta _{k}}  \label{squeezed state}
\end{equation}%
where
\begin{eqnarray}
\beta _{k} &=&\alpha +(-1)^{k}\frac{i\eta }{\Omega _{k}N_{k}}\left(
1-e^{i\Omega _{k}t}\right)  \nonumber \\
\theta _{k} &=&\left( \frac{\eta ^{2}}{\Omega _{k}N_{k}}+\left( -1\right)
^{k}E_{J}\cos \phi _{e}\right) t \\
\mu _{k} &=&\cos \Omega _{k}t+i\frac{\omega +\left( -1\right) ^{k}2\delta }{%
\Omega _{k}}\sin \Omega _{k}t \\
\nu _{k} &=&-\left( -1\right) ^{k}i\frac{2\delta }{\Omega _{k}}\sin \Omega
_{k}t  \nonumber \\
\Omega _{k} &=&\sqrt{\omega ^{2}+(-1)^{k}4\delta \omega } \\
N_{k} &=&\frac{\sqrt{\omega ^{2}+(-1)^{k}4\delta \omega }}{\omega }
\end{eqnarray}%
and $\beta _{k}=\alpha $ when the time $t=0$, for $k=0,1$ respectively. Here
the squeezed states $\left\vert \beta _{k},\mu _{k},\nu _{k}\right\rangle $
is defines as in \cite{Scully97} for a new set of boson operators \cite%
{Wang02}
\[
A_{k}=\mu _{k}a-\nu _{k}a^{\dagger }
\]%
(for $k=0,1$) which hold
\[
A_{k}\left\vert \beta _{k},\mu _{k},\nu _{k}\right\rangle =\beta
_{k}\left\vert \beta _{k},\mu _{k},\nu _{k}\right\rangle .
\]

Then the time evolution of decoherence factor which characters the quantum
coherence of the charge qubit is
\begin{eqnarray}
D\left( t\right) &=&\frac{1}{\sqrt{\mu _{1}\mu _{0}^{\ast }-\nu _{1}\nu
_{0}^{\ast }}} \\
&&\exp \left\{
\begin{array}{c}
-\frac{1}{2}\left\vert \beta _{0}\right\vert ^{2}-\frac{1}{2}\left\vert
\beta _{1}\right\vert ^{2} \\
+\frac{2\beta _{1}^{\ast }\beta _{0}+\beta _{0}^{2}\left( \mu _{1}^{\ast
}\nu _{0}-\nu _{1}^{\ast }\mu _{0}\right) +\beta _{1}^{\ast 2}\left( \nu
_{1}\mu _{0}-\nu _{0}\mu _{1}\right) }{2\left( \mu _{1}^{\ast }\mu _{0}-\nu
_{1}^{\ast }\nu _{0}\right) }%
\end{array}
\right\} .  \nonumber
\end{eqnarray}

\begin{figure}[ht]
\includegraphics[width=8cm]{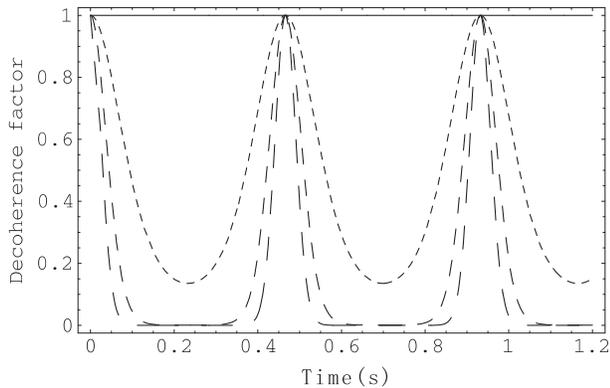}
\caption{Time evolution of the decoherence factor $D\left( t\right) $ with $%
\protect\phi _{e}=0$ and different $\protect\alpha =0,1,2,3$ respectively (
from top to down). When $\protect\alpha \neq 0$, the second order
interaction in Eq.(\protect\ref{Ham6}) induces collapses and revivals of
quantum coherence. When $\protect\alpha =0$, the qubit owns the least loss
of quantum coherence.}
\label{t0}
\end{figure}

\begin{figure}[ht]
\includegraphics[width=8cm]{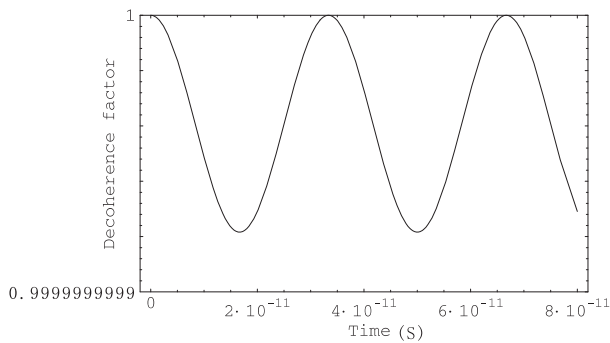}
\caption{Time evolution of the decoherence factor $D\left( t\right) $ with $%
\protect\phi _{e}=\frac{\protect\pi }{2}$. It means that the qubit owns the
least loss of quantum coherence.}
\label{t3}
\end{figure}

We control the interaction between qubit and cavity by changing the strength
of the external classical field, i.e., . From Eq.(\ref{Ham6}), if we turn
off the external field, i.e., $\sin \phi _{e}=0$, the above Hamiltonian only
includes the second order interaction. If we turn on the external magnetic
field, i.e., $\cos \phi _{e}=0$, the above Hamiltonian only includes the
first order interaction.

Typically in Fig.\ref{t0} and Fig.\ref{t3}, the time evolution of
decoherence factor $D\left( t\right) $ demonstrates an engineered
decoherence process. We find that the very sharp peaks in $D(t)$ curves may
originate from the reversibility of the Schr\"{o}dinger equation for few
body system and we called them quantum jumps \cite{Sun93}, i.e., the quantum
coherence for charge qubit exhibits collapses and revivals. In Fig.\ref%
{cavity}, when the interaction Hamiltonian only includes the second order
interaction, i.e., at the value of $\phi _{e}$ for $\sin \phi _{e}=0$, the
period of oscillation for $D\left( t\right) $ is unchanged and the width of
the peak becomes smaller with $\alpha =1,2,3$ respectively. Specially at the
value of $\phi _{e}$ for $\sin \phi _{e}=0$ and $\alpha =0$, $D\left(
t\right) $ approaches one, i.e., the loss of coherence is negligible. In Fig.%
\ref{t3}, when$\ $the interaction Hamiltonian only includes the first order
interaction, i.e., cos$\phi _{e}=0$, we have the time evolution of the
decoherence factor
\begin{equation}
D\left( t\right) =\exp \left[ \frac{8\eta ^{2}}{\omega ^{2}}\sin ^{2}\frac{
\omega t}{2}\right] .
\end{equation}
It means that $D\left( t\right) $ is independent of values of $\alpha $\ and
the amplitude of oscillation is very small.

\section{discussions}

In this paper, we have studied quantum entanglement in the charge
qubit-cavity system and demonstrate a engineered decoherence process in the
case of the second order approximation. In Fig.\ref{t0} and Fig.\ref{t3}, we
find that controlled parameter $\left( \phi _{e},\alpha \right) $ exists two
optimal values $\left( \sin \phi _{e}=0,\alpha =0\right) $ and $\left( \cos
\phi _{e}=0,\alpha \right) $ with the least loss of quantum coherence for
the charge qubit.

On the other hand, we demonstrate a dynamical process of quantum
information- storage between the charge qubit and the microwave cavity. From
the above results, we get the time of the information storage $\Delta t=
\frac{\pi }{2\eta }\sim \frac{\sqrt{V}}{S}$. To implement faster operation
of information storage, we can reduce the electromagnetic mode volume $V$ of
microwave cavity and increase the bounded area $S$ of dc SQUID (charge
qubit). We wish that our results in this paper will be helpful to
experiments.

\begin{acknowledgments}
We thank the support of the CNSF (grant No.90203018) and the
Knowledged Innovation Program (KIP) of the Chinese Academy of
Sciences and the National Fundamental Research Program of China
with No.001GB309310. We also sincerely thank Y. Li, P. Zhang and
C. P. Sun for helpful discussions.
\end{acknowledgments}


\begin{thebibliography}{DiVincenzo(2000)}
\bibitem[Zurek(1983)]{Zurek83} J. A. Wheeler and Z. H. Zurek, \textit{\
Quantum Theory of Measurement} (Princeton University Press, NJ, 1983).

\bibitem[Makhlin(2001)]{Makhlin01} Y. Makhlin, G. Schon, and A. Shnirman,
Rev. Mod. Phys. \textbf{73} (2001) 357 .

\bibitem[Nakamura(1999)]{Nakamura99} Y. Nakamura, Yu. A. Pashkin, and J.S.
Tsai, Nature (London) \textbf{398} (1999) 786.

\bibitem[Pashkin(2003)]{Pashkin03} Yu. A. Pashkin, T. Yamamoto, O. Astafiev,
Y. Nakamura, D. V. Averin and J. S. Tsai, Nature (London) 421 (1999) 823.

\bibitem[Han(2002)]{Han02} Y. Yu, S. Han, X. Chu, Shih-I Chu and Z. Wang,
Science 296 (2002) 889.

\bibitem[Raimond(2001)]{Raimond01} J. Raimond, M. Brune, S. Haroche, Rev.
Mod. Phys. \textbf{73} (2001) 565.

\bibitem[Lukin(2000)]{Lukin00} M.D. Lukin, S.F. Yelin, and M.
Fleischhauer,Phys. Rev. Lett. \textbf{84} (2000) 4232 .

\bibitem[Lukin(2003)]{Lukin03} M.D. Lukin, Rev. Mod. Phys. \textbf{75}
(2003) 457.

\bibitem[Li(2003)]{Li03} Y. Li and C.P. Sun, LANL eprint, quant-ph/0312093.

\bibitem[Cleland(2003)]{Cleland03} A. N. Cleland and M. R. Geller, LANL
eprint, cond-mat/0311007.

\bibitem[Han(2003)]{Han03} C.-P. Yang, S.-I. Chu and S. Han, Phys. Rev.
\textbf{A67} (2003) 042311.

\bibitem[You(2003)]{You03} J.Q. You and Franco Nori, Physica E \textbf{18}
(2003) 33-34.

\bibitem[Girvin(2003)]{Girvin03} S. M. Girvin, Ren-Shou Huang, Alexandre
Blais, Andreas Wallraff and R. J. Schoelkopf, LANL eprint, cond-mat/0310670.

\bibitem[Nakamura(2002)]{Nakamura02} Y. Nakamura, Yu.A. Pashkin, T. Yamamoto
and J.S. Tsai, Phys. Rev. Lett. \textbf{88} (2002) 047901.

\bibitem[Makhlin(1999)]{Makhlin99} Y. Makhlin, G. Schon, and A. Shnirman,
Nature (London) \textbf{386} (1999) 305 .

\bibitem[Scully(2003)]{Scully97} M.O. Scully and M.S. Zubairy, \textit{\
Quantum Optics} (Cambridge, England, 1997).

\bibitem[DiVincenzo(2000)]{DiVincenzo00} D. DiVincenzo, Fortschr. Phys.
\textbf{48} (2000) 771 .

\bibitem[Jaynes(1963)]{Jaynes63} E.T. Jaynes, F.W. Cummings, Proc. IEEE
\textbf{51} (1963) 89 .

\bibitem[Raimond(1997)]{Raimond97} J.M. Raimond, M. Brune and S. Haroche,
Phys. Rev. Lett. \textbf{79} (1997) 1964 .

\bibitem[Sun(2000)]{Sun00} H. B. Zhu, C. P. Sun, Chinese Science (A) 2000.10
\textbf{30}(10) 928-933; Progress in Chinese Science, 2000.60 \textbf{10}(8)
698-703.

\bibitem[Wang(2002)]{Wang02} Y.D. Wang, Y.B. Gao and C.P. Sun, LANL eprint,
quant-ph/0212088.

\bibitem[Sun(1993)]{Sun93} Chang-Pu Sun, Phys. Rev. \textbf{A48} (1993) 898
. C. P. Sun et.al, Fortschr. Phys. \textbf{43} (1995) 585 .
\end{thebibliography}
\end{document}